\documentclass[sigconf,authorversion,nonacm]{aamas} 

\usepackage{balance} 
\usepackage{diagbox} 
\usepackage{hyperref} 
\usepackage[noabbrev,nameinlink,capitalize]{cleveref} 
\usepackage{amsmath} 

\usepackage{amssymb} 
\usepackage{siunitx} 

\crefname{subsection}{Subsection}{Subsections}

\title[Emergent Dominance Hierarchies in Reinforcement Learning Agents]{Emergent Dominance Hierarchies \\ in Reinforcement Learning Agents}
\titlenote{A condensed version of this work has been presented as an extended abstract at the 2024 International Conference on Autonomous Agents and Multiagent Systems (AAMAS) \citep{rachum2024dh-ea}}

\author{Ram Rachum}
\affiliation{
  \institution{Independent Researcher}
}
\email{ram@rachum.com}

\author{Yonatan Nakar}
\affiliation{
  \institution{Tel Aviv University}
}

\author{Bill Tomlinson}
\affiliation{
  \institution{University of California, Irvine}
}

\author{Nitay Alon}
\affiliation{
  \institution{Max Planck Institute, Bio-Cybernetics}
}
\affiliation{
  \institution{Hebrew University of Jerusalem}
}

\author{Reuth Mirsky}
\affiliation{
  \institution{Bar-Ilan University}
  \country{Israel}
  }

\begin{abstract}

    Modern Reinforcement Learning (RL) algorithms are able to outperform humans in a wide variety of tasks. Multi-agent reinforcement learning (MARL) settings present additional challenges, and successful cooperation in mixed-motive groups of agents depends on a delicate balancing act between individual and group objectives. Social conventions and norms, often inspired by human institutions, are used as tools for striking this balance.

    In this paper, we examine a fundamental, well-studied social convention that underlies cooperation in both animal and human societies: dominance hierarchies.

    We adapt the ethological theory of dominance hierarchies to artificial agents, borrowing the established terminology and definitions with as few amendments as possible. We demonstrate that populations of RL agents, operating without explicit programming or intrinsic rewards, can invent, learn, enforce, and transmit a dominance hierarchy to new populations. The dominance hierarchies that emerge have a similar structure to those studied in chickens, mice, fish, and other species.

\end{abstract}


\newcommand{\BibTeX}{\rm B\kern-.05em{\sc i\kern-.025em b}\kern-.08em\TeX}

\begin{document}


\pagestyle{fancy}
\fancyhead{}

\maketitle

\section{Introduction}

Many animal species are able to collaborate, form groups and harness collective intelligence \citep{couzin2009collective, gordon1996ants}. Among those species, humans have achieved a scale and sophistication of collaboration that stands as one of the most profound and unparalleled phenomena on Earth. The grand feats of humanity were accomplished by groups of humans working together; what is grander is that these groups are composed of individuals with wildly different beliefs, motivations, and talents \citep{woolley2010evidence}. While differences between individuals provide an important source of diversity, they also lead to adversity and conflict, which can result in the violent demise of the social group \citep{zhan2015sword}. Those human civilizations that tempered their internal conflicts while capitalizing on the diversity were the ones to survive and flourish \citep{boyd2009culture, bowles2009warfare, alesina2005ethnic}. The intricate social structures that make up these civilizations turn a potential runaway explosion into a controlled reaction, enabling groups of humans to achieve breakthroughs that exceed the combined capabilities of their individual members \citep{muthukrishna2016innovation}.

Multi-agent reinforcement learning (MARL) presents an opportunity to implement simplified versions of those intricate social structures, where the environment is closed and controlled, and humans are replaced by artificial agents. Research in the paradigm of Cooperative AI \citep{dafoe2020caif, dafoe2021nature, conitzer2023foundations, jaques2019social, vinitsky2023sanctions, radke2023better} highlights the role of such social structures (also labeled \textit{institutions}) in producing human-aligned AI systems. Existing research into reproducing such institutions for use in MARL settings includes language \citep{mordatch2018emergence}, voting systems \citep{conitzer2006voting, xu2020votingbased}, auctions \citep{parsons2011auctions}, reputation systems \citep{ramchurn2009trust}, and bargaining \citep{kraus1997}.
 
In this paper we take a step back from these cerebral and centralized institutions and turn our attention to a primordial institution that underlies cooperation in both animal and human societies: \textit{dominance hierarchies}.

Our contribution in this work is twofold:

First, we adapt the ethological theory of dominance hierarchies to artificial agents by modeling conflicts between each pair of individuals as the classic game of Chicken \citep{rapoport1966chicken}.\footnote{We use the word ``chicken'' in this paper to refer to both the Game Theory concept and the beak-and-feathers animal. We capitalize ``Chicken'' to indicate the former.} To accommodate groups of more than two individuals, we generalize the game of Chicken into an $N$-player stochastic game that we call \textit{Chicken Coop}. We borrow the established terminology and definitions from animal study to allow calculating dominance metrics from agents' action history. We release the Chicken Coop environment under the MIT open-source license.\footnote{The code and installation instructions are available at \url{https://github.com/cool-RR/chicken-coop}}

Second, we train RL agents on the Chicken Coop environment to optimize their score, observing the emergence of three behaviors:

\begin{enumerate}
    \item Agents collaboratively invent dominance hierarchies.
    \item Agents enforce dominance hierarchies on other agents.
    \item Agents transmit dominance hierarchies to new populations.
\end{enumerate}

These behaviors emerge from minimal environment rules and individual behavior determined by a stock RL algorithm with no intrinsic rewards.

We experiment with gradually ablating the agents' observations, showing a causal link between the agents' ability to identify the opposing agent and the emergence of dominance hierarchies. We compare the attributes of dominance hierarchies that emerge to empirical results from groups of different animals, showing structural similarities in both aggressiveness levels and the distribution of intransitive triads.

This project seeks to lay the intellectual groundwork for dominance hierarchies to feature prominently in future MARL systems. Dominance hierarchies have been shown to provide substantial value to both animal and human communities \citep{mintzberg1989structuring, smith1973logic}. By providing MARL systems with the ability to coordinate their collective behavior via dominance hierarchies, we hope to enable those systems to reap similar benefits and integrate more seamlessly into existing human systems \citep{reeves1996media}.

\section{Background}

\subsection{Dominance hierarchies in nature}

When several animals live together in a social group they often find themselves in conflicts with other members of the group over resources such as food or mating partners. These conflicts can lead to physical injury and death. A \textit{dominance relationship} \citep{rapoport1949outline} is a pattern in the conflicts between a pair of individuals, in which the \textit{dominant} individual is highly likely to escalate violence and win the majority of resources, and the \textit{subordinate} individual is highly likely to deescalate and yield any contested resources. This dynamic prevents the runaway escalation of violence that could otherwise lead to individual injury and a loss of group cohesion \citep{smith1973logic}. 

A \textit{dominance hierarchy}, also known as a \textit{pecking order}, is the aggregate of all the dominance relationships between each pair of agents in the group. It is defined by modelling dominance relationships as a total relation on the set of agents, where $d \rightarrow s$ means that individual $d$ is dominant over individual $s$. A dominance hierarchy is modelled as a complete, directed graph (tournament \citep{diestel2017graph}) where agents are represented as nodes and dominance relationships are represented as directed edges \citep{iverson1990statistical}. This graph representation is more than just a map of agonistic behavior in the group; any skirmish between individuals $a$ and $b$ affects not only the edge connecting $a$ and $b$, but the edges connecting either $a$ or $b$ with each of the other individuals in the group, by the mechanisms of \textit{winner effects} \citep{chase1994aggressive}, \textit{loser effects} \citep{dugatkin1997winnerloser} and \textit{bystander effects} \citep{dugatkin2001bystander}.

The field of dominance hierarchies traces its roots to \citet{schjelderup1922beitrage}, which described pecking orders in captive chicken societies. Over the past century, dominance hierarchies have been studied across a wide range of animal species including canines \citep{bonanni2017age,essler2016play}, birds \citep{watson1970dominance}, elephants \citep{wittemyer2007hierarchical}, fish \citep{david2003spatial} and primates \citep{goodall1986chimpanzees, sapolsky2005influence}, uncovering commonalities in the structure and function of dominance hierarchies across taxa \citep{chase2022networks}. The study of dominance hierarchies continues to be a field of active research \citep{strauss2022dynamics}.

\subsection{Dominance hierarchies in humans}

In human societies, dominance hierarchies appear in many forms, ranging from explicit and rigid to implicit and fluid \citep{magee2008social, halevy2012status}. The explicit forms, often appearing in corporations, military units, and sports teams, are laid bare for analysis, as they are announced with official titles, insignia patches, and organizational charts. The ``chain of command'' principle that underpins large-scale human enterprise is an extension of human dominance behavior to groups that are far too large for any single individual to comprehend \citep{mintzberg1989structuring, grueneisen2017children, von2019tsimane}.

Implicit forms of dominance hierarchies often go unspoken, making them more difficult to study; however, through subtlety, their effect on human behavior is profound \citep{anderson2009pursuit, tiedens2003power}. \citet{Johnstone2007Impro} provides a manual for theater actors to faithfully reproduce real-world social scenarios, noting the importance of dominance hierarchies:

\begin{quote}
``Try to get your status just a little above or below your partner's,'' I said, and I insisted that the gap should be minimal. The actors seemed to know exactly what I meant and the work was transformed. The scenes became `authentic', and actors seemed marvellously observant. Suddenly we understood that every inflection and movement implies a status [\ldots] It was hysterically funny, but at the same time very alarming. All our secret manoeuvrings were exposed. [\ldots] Normally we are `forbidden' to see status transactions except when there's a conflict. In reality status transactions continue all the time.
\end{quote}

We note that, while dominance hierarchies form an important part of human social interactions, they are nevertheless just one component of the complex array of motivations that together produce social behavior \citep{chenzeng2022dominance}.

\section{Related Work}

While we advocate using dominance hierarchies to improve the performance of artificial agents, considerable work has been done in the opposite direction, applying MAS and MARL to improve our understanding of dominance hierarchies in animals. We use this distinction to roughly divide the work in the intersection of these fields to two categories, detailed below.

\subsection{Applying MAS/MARL to dominance hierarchies}

MAS models have been used for the modelling of dominance hierarchies in animals \citep{ishiwaka2022deepfoids, gavrilets2008dynamics} and in humans \citep{przepiorka2020dominance}. \textit{DomWorld} \citep{hemelrijk2000domworld} is a multi-agent environment developed to study the self-organization of social behavior and dominance hierarchies in primate groups. DomWorld demonstrates the decentralized nature of dominance hierarchies: each agent's actions are determined according to its own logic and observations, and the collective actions of all agents cause dominance hierarchies to emerge on the population level. These dominance hierarchies have similar properties to those found in real animal societies. Notably, DomWorld is not modelled as a stochastic game or any other game theoretic construct, as the agents have no concept of a reward signal that they're trying to maximize. Agents follow hard-coded logic to decide whether to attack an opponent or flee, helpfully described in a flowchart in \citet{hemelrijk2017domworld}.

\citet{leimar2021evolution} provides a proper MARL environment called \textit{SocDom}, equipping the agents with an RL algorithm \citep{sutton2018reinforcement}, albeit without neural network estimators. While the SocDom experiments show similar results to ours, the foci of the two environments differ: SocDom, like DomWorld, attempts to tailor the agents' behaviors to faithfully reproduce the dominance behavior seen in real animal societies. This is evident in SocDom's use of genetic algorithms in addition to RL. In contrast, our research is directed at boiling the mechanics of dominance hierarchies down to the simplest possible set of premises, showing that they emerge even when using an off-the-shelf RL algorithm.

\subsection{Applying dominance hierarchies to MAS/MARL}

The use of dominance hierarchies for improving the performance of artificial agents was proposed in \citet{hemelrijk1996dominance}, and in \citet{tomlinson2000wolf}, which suggested that dominance hierarchies could streamline negotiation, enable agents to form beneficial alliances, and make interfaces more intuitive.

\citet{bakker2021modelling} employs a group of RL agents with a graph structure, which bears resemblance to a dominance hierarchy, besides being explicitly defined rather than emergent. The authors show how the dominance-like relationship between each pair of individuals results in increased cooperation from both sides.

\citet{aroca2023hierarchical} demonstrates an explicit hierarchy where some agents have a different set of actions than others, depending on their position in the hierarchy. The authors show that the hierarchy improves the ability of humans to interact with the system, as humans find the agents more fluent and trustworthy.

\section{Definitions}

In this section we adapt the terms and definitions used in the study of dominance hierarchies in animal societies to the frameworks used in multi-agent reinforcement learning.

\subsection{Dominance between two agents}
\label{subsec:definitions_two_agents}

Dominance hierarchies are a group phenomenon comprised of multiple dominance relationships, one between each pair of agents in the group. In this subsection we consider only those pairwise interactions; we define dominance relationships, the metrics of aggressiveness and rapport, the roles of dominant and subordinate, and the environment in which these occur.

Animals use dominance relationships to decide which individual will have access to an exclusive resource, such as food, a mating partner, or grooming by other individuals \citep{henazi1999grooming}. We model the environment that enables these interactions between agent $i$ and agent $j$ as a partially-observable stochastic game (POSG) \citep{hansen2004dynamic, yang2020overview} which has exactly two stable \citep{kohlberg1986strategic} Nash equilibria $NE_i$ and $NE_j$ such that agent $i$'s reward at $NE_i$ is bigger than its reward at $NE_j$, and agent $j$'s reward at $NE_j$ is bigger than its reward at $NE_i$. 

We denote the actions that comprise $NE_i$  as $a_i^{NE_i}$ and $a_j^{NE_i}$.

Intuitively, a dominance relationship between two agents is the tendency of those agents to play the joint action of either $NE_i$ or $NE_j$, but not both. Similarly to \citet{leibo2017ssd}, we define an agent's aggressiveness by how frequently it chooses the action that reduces the other agent's reward:

\begin{definition}[Aggressiveness]

    Given an agent $i$ and a set of timesteps $\mathbb{T}$, the agent's \textit{aggressiveness} $g_i^\mathbb{T}$ is the portion of timesteps in which it played $a_i^{NE_i}$:
    \[
        g_i^\mathbb{T}=\frac{|\{t \in \mathbb{T} \mid a_i^t = a_i^{NE_i}\}|}{|\mathbb{T}|}
    \]

\end{definition}

\begin{definition}[Dominance relationship, dominant, subordinate]
    \label{def:dominance}
    Given two agents $i$ and $j$ and a set of timesteps $\mathbb{T}$ in which they played with each other, if the difference between agent $i$'s aggressiveness and agent $j$'s aggressiveness is above a certain threshold, we say that the two agents are in a \textit{dominance relationship} (DR), with agent $i$ being \textit{dominant} and agent $j$ being \textit{subordinate}:

    \begin{align*}
        i \rightarrow j \quad \text{iff} \quad g_i^\mathbb{T} - g_j^\mathbb{T} > \eta, \quad \eta \in (0, 1]
    \end{align*}
\end{definition}

In order to determine the existence of a dominance relationship between two agents without making assumptions about its polarity, we introduce a metric we call \textit{rapport}:

\begin{definition}[Rapport]
    
    For two agents $i$ and $j$ and timesteps $\mathbb{T}$, we define the \textit{rapport} $\mathfrak{R}_{i,j}^\mathbb{T}$ as:

    \[\mathfrak{R}_{i,j}^\mathbb{T} = \lvert g_i^\mathbb{T} - g_j^\mathbb{T}\rvert\]
    
\end{definition}

A rapport $\mathfrak{R}_{i,j}^\mathbb{T} > \eta$ indicates either $i \rightarrow j$ or $i \leftarrow j$.

There are infinitely many environments that satisfy the conditions needed for dominance relationships to arise. \citet{przepiorka2020dominance} defines the Hierarchy Contest Game (HCG), which is a generalization of the classic game of Chicken \citep{rapoport1966chicken} to a game where players have different capabilities. They show how dominance hierarchies form in human play, whether the players start with knowledge of each other's capabilities or not. Because we wish to describe the simplest possible environment that can produce dominance relationships, we use the game of Chicken with no modifications. Its salient feature is the pair of symmetrically opposing, stable, pure-strategy Nash equilibria.

In the game of Chicken, each of the two agents plays either \textit{hawk} or \textit{dove}. An agent gets the most reward if it plays $hawk$ while its partner plays $dove$; however, if both agents play $hawk$, they both get the lowest reward. 

\begin{definition}[The game of Chicken]
  The game of Chicken is defined as a normal-form game:    
  \begin{center}    
    \begin{tabular}{|c|c|c|}
      \hline
      \diagbox{Agent 1}{Agent 2} & Dove & Hawk \\
      \hline
      Dove & $R, R$ & $S, T$ \\
      \hline
      Hawk & $T, S$ & $P, P$ \\
      \hline
    \end{tabular}
  \end{center}
  The reward constants satisfy $T > R > S > P$. The $(S, T)$ and $(T, S)$ outcomes serve as the Nash equilibria $NE_i$  and $NE_j$ . Each agent's aggressiveness is simply its tendency to play \textit{hawk}.
\end{definition}

\subsection{Dominance hierarchies}

\begin{figure}[t]
    \includegraphics[width=\linewidth]{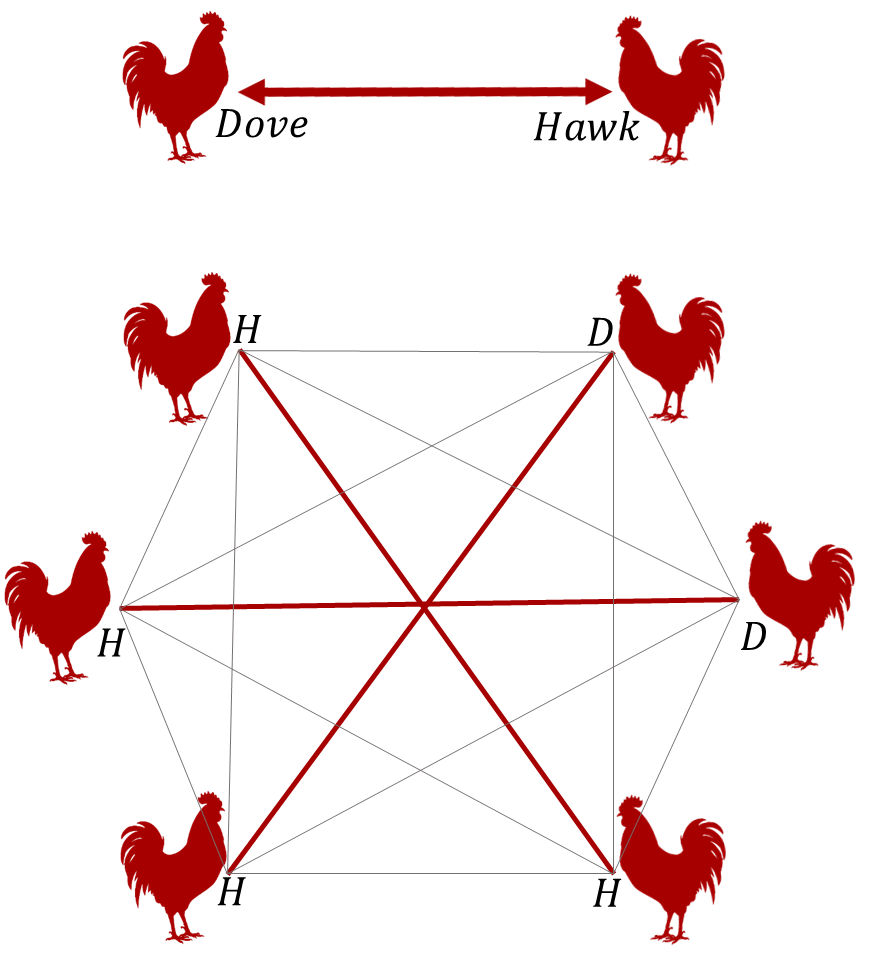}
    \caption{The game of Chicken (\textbf{top}) compared with Chicken Coop (\textbf{bottom}). Each Chicken Coop episode can be described as $N/2$ chicken games between randomly-paired agents.}
    \label{fig:coop_pictogram}
\end{figure}

To model dominance relationships within an agent population, we generalize the game of Chicken to support more than two players, visualized with $N=6$ in \cref{fig:coop_pictogram}:

\begin{definition}[Chicken Coop]
    \textit{Chicken Coop} is an $N$-player generalization of the game of Chicken. In each episode of Chicken Coop, the agents are divided into random pairs that are used for the entire episode. Each pair of agents plays one round of Chicken against each other, choosing either \textit{hawk} or \textit{dove} and receiving a reward in $\{T, R, S, P\}$. Each agent's sole observation is the identity of their opponent agent.
\end{definition}

We extend the definitions in \cref{subsec:definitions_two_agents} from the Chicken game to the Chicken Coop game. For example, for each pair of agents $i$ and $j$ in the population, we determine the existence and polarity of their dominance relationship ($i \rightarrow j$ or $i \leftarrow j$) by evaluating the aggressiveness metrics $g_i^\mathbb{T}$ and $g_j^\mathbb{T}$ on the subset of $\mathbb{T}$ in which these two agents were randomly paired with each other. Finally, we aggregate these dominance relationships into a graph:

\begin{definition}[Dominance hierarchy]

    A \textit{dominance hierarchy} $\mathcal{H}$ is a complete, directed graph (also known as a tournament \citep{diestel2017graph},) where agents are represented as nodes and dominance relationships are represented as directed edges.

\end{definition}

\section{Methods}

In all of the experiments below, unless stated otherwise, the RL agents are trained using the PPO algorithm \citep{Schulman2017Proximal} implementation provided by the RLlib framework \citep{liang2018rllib}. We train $L = 300$ populations of $N=6$ Chicken Coop agents, at a learning rate of $2 \times 10^{-6}$, a discount factor of $\gamma = 0.99$, and a clipping parameter of $\epsilon = 0.3$. We use a dominance threshold of $\eta = 0.55$ and these reward constants: 

\[
R = 0, \quad S = -2, \quad T = 5, \quad P = -10
\]

Each agent has its own neural network with its own set of weights, which are separate from those of the other agents. Crucially, all of the agents in a given population are trained exclusively with the other agents in that population; we don't allow cross-population training, because we would like to allow each population to develop its own distinct dominance hierarchy.

Each generation is comprised of $512$ episodes. After each generation, the policies update according to the actions and rewards in that generation.

\section{Results}
\label{sec:results}

\subsection{Emergence of dominance hierarchies}
\label{subsec:emergence}

\begin{figure}[t]
    \includegraphics[width=\linewidth]{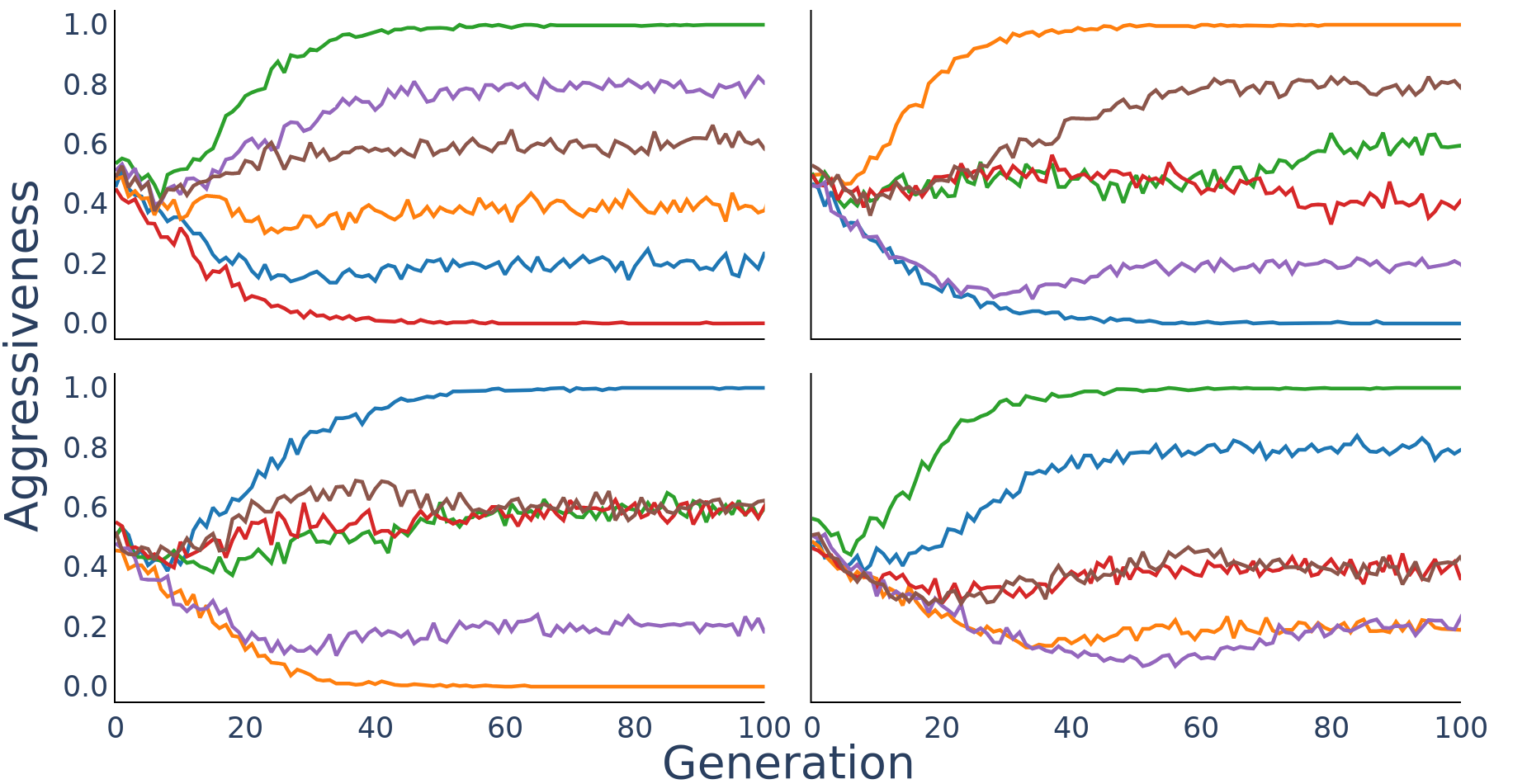}
    \caption{Aggressiveness levels of each of the six agents (depicted as differently-colored lines) in four sample Chicken Coop populations.}
    \label{fig:samples_aggressiveness}
\end{figure}

\begin{figure}[t]
    \includegraphics[width=\linewidth]{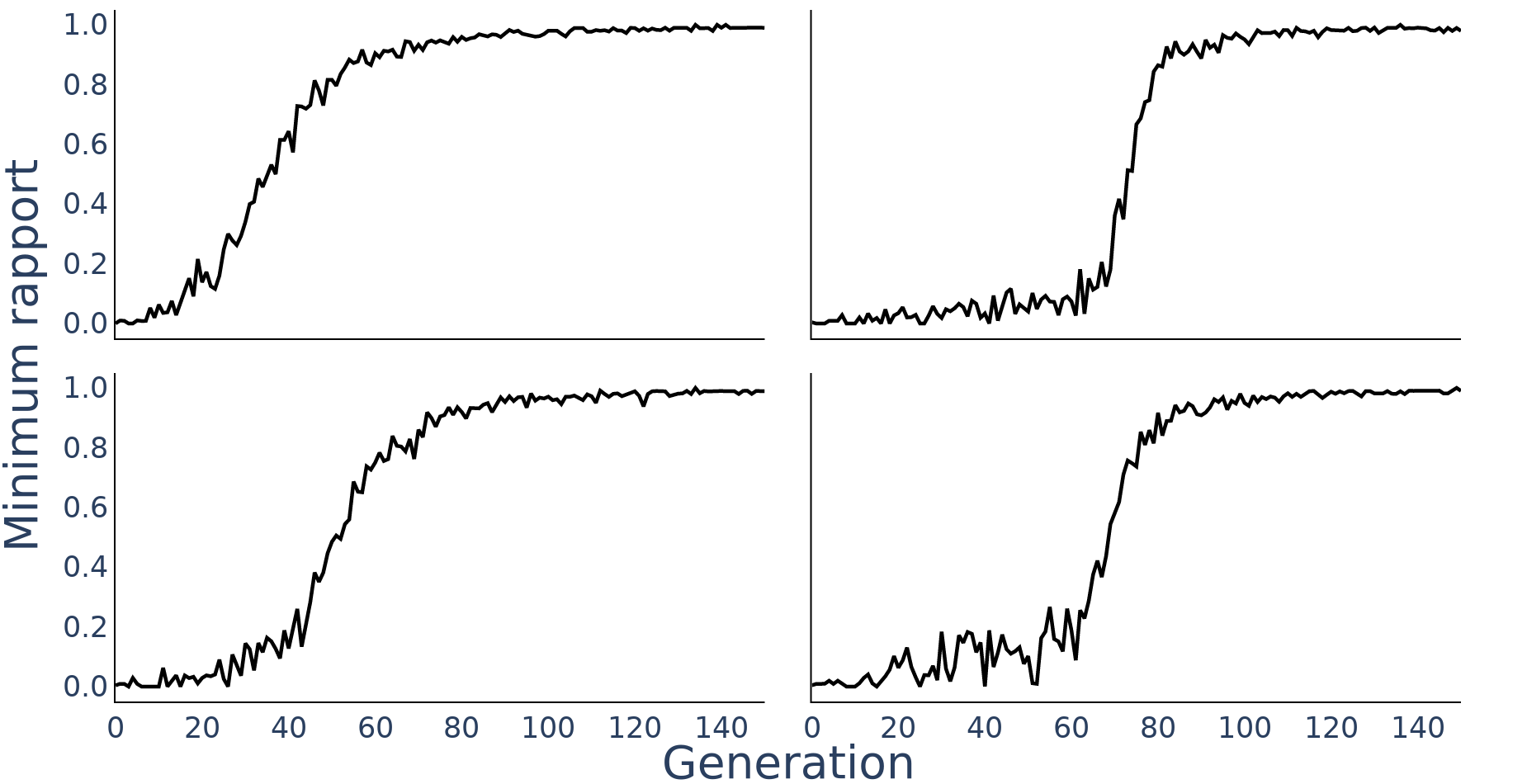}
    \caption{Minimum rapport between each pair of the six agents in four sample Chicken Coop populations.}
    \label{fig:samples_rapport}
\end{figure}

We measure the aggressiveness of each agent in each population, aggregated from its interactions with each of the five other agents. In \cref{fig:samples_aggressiveness} we plot the aggressiveness of all six agents in four sample populations. A few features can be visually discerned:
\begin{enumerate}
    \item The agents' aggressiveness values converge to approximately equally-spaced ``ranks'' in $[0, 1]$.
    \item Despite of this strong convergence, the mapping between agents and ranks appears completely arbitrary across the four sample populations.
    \item In some populations, the mapping between agents and ranks is a bijection, while in other populations, multiple agents converge to the same rank, leaving other ranks empty.
\end{enumerate}

We offer the following explanation for these observations: when agent $i$'s aggressiveness approaches $k/N$, agent $i$ is playing a high proportion of \textit{hawk} towards $k$ other agents, which in turn play a high proportion of \textit{dove} towards agent $i$. Each of these $k$ agents contributes $1/N$ to agent $i$'s aggressiveness. According to \cref{def:dominance}, agent $i$ is dominating these $k$ agents. The remaining $N-k-1$ agents dominate agent $i$.

We verify this explanation by measuring the minimum rapport between each pair of agents for each of the four sample populations, observing that each agent develops a dominance relationship with each of the other agents in its population. (\cref{fig:samples_rapport}.)

\begin{figure}[t]
    \includegraphics[width=\linewidth]{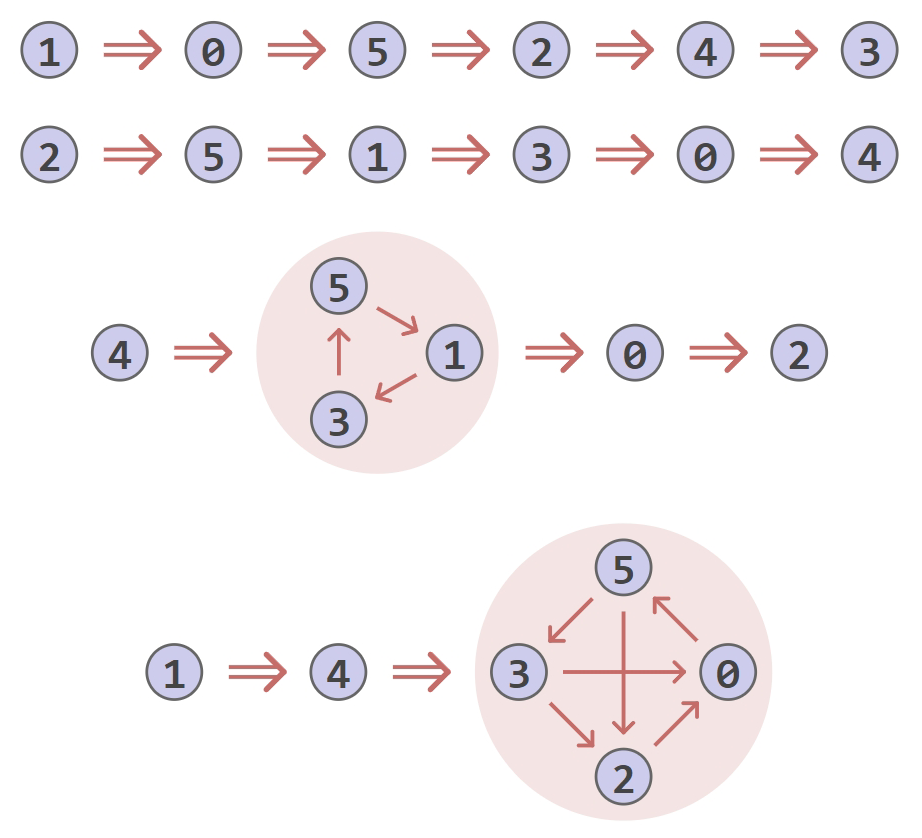}
    \caption{Dominance hierarchies of four sample Chicken Coop populations, visualized in condensed form. A single arrow ($\rightarrow$) represents a dominance relationship between two agents. A double arrow ($\Rightarrow$) represents a transitive relationship, i.e. each of the agents on the left side of the double arrow dominates each of the agents on the right side of that arrow.}
    \label{fig:samples_dh}
\end{figure}

We aggregate these dominance relationships into dominance hierarchies, showing the results for our four sample populations in \cref{fig:samples_dh}. We observe that some of the dominance hierarchies form a perfect line, while others contain cycles. This distinction has been studied in animal societies. The former are called \textit{linear dominance hierarchies} or \textit{transitive dominance hierarchies}. The latter are called \textit{nonlinear dominance hierarchies}, \textit{near-linear dominance hierarchies} or \textit{intransitive dominance hierarchies} \citep{chase2002individual, devries1995improved, devries1998finding, sanders2003resource}. In linear dominance hierarchies, the agents may be assigned ranks $\{0, 1, 2, \ldots, N-1\}$ with rank $0$ representing the agent that dominates all other agents, rank $1$ representing the agent that dominates all other agents except that in rank $0$, etc.

One profound property of the agents' learned behavior is its idiosyncraticity: (1) it converges to a stable dominance hierarchy, and (2) this hierarchy is not a single one, but rather varies between different trials. Our $300$ populations converged to a total of $248$ distinct dominance hierarchies (out of $2\frac{N(N-1)}{2} = 2^{15}$ possible permutations). Most dominance hierarchies appeared in only one population, while the most common dominance hierarchies appeared in $3$ populations. Despite the arbitrariness of that choice, each population converges strongly to its chosen dominance hierarchy. Out of $248$ distinct dominance hierarchies, only $17$ were intransitive, i.e. containing dominance cycles.

\subsection{Comparison to animal behavior}
\label{subsec:comparison_animal}

\begin{figure}[t]
    \includegraphics[width=\linewidth]{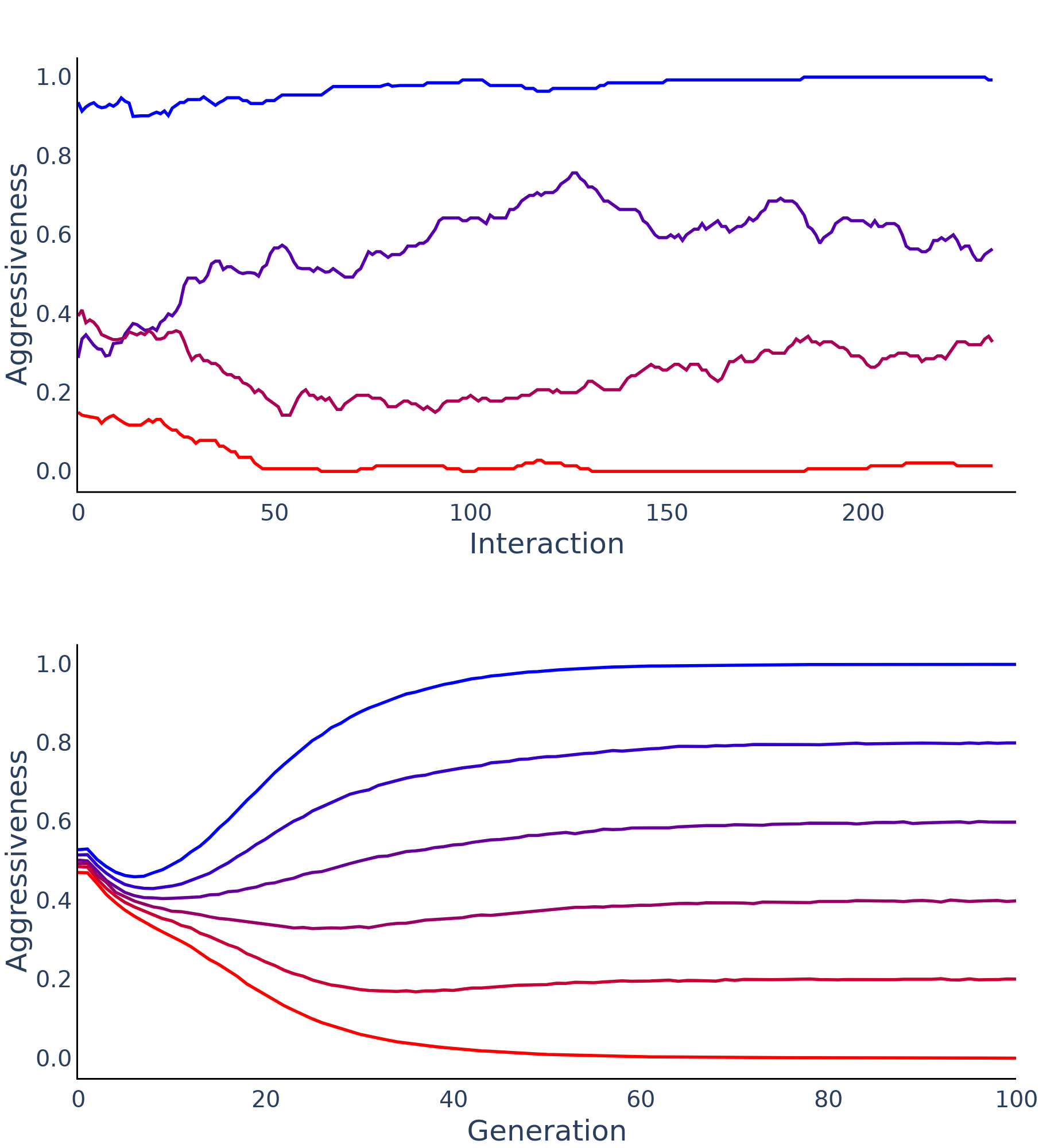}
    \caption{\textbf{Top:} Mean aggressiveness levels of each of four Leghorn chickens in fourteen captive populations \citep{chase2022networks}. \\\textbf{Bottom:} Mean aggressiveness levels of each of six Chicken Coop agents across 300 populations, filtered to include only those that developed a linear dominance hierarchy. \\ In both plots, the results are averaged using the agents' ranks as their identity.}
    \label{fig:comb_hero}
\end{figure}

In \cref{fig:comb_hero} we visually compare our aggressiveness results to empirical data from groups of actual chickens. In \citet{chase2022networks}, 14 captive populations of 4 white Leghorn hens each were observed for aggressive behavior for two consecutive days. While there are considerable differences in the methodologies of that experiment and our study, the property of idiosyncraticity holds true in both of them. In both environments, the agents tend to maintain their position, regardless of how high or low it is in the hierarchy.

\begin{figure}[t]
    \includegraphics[width=\linewidth]{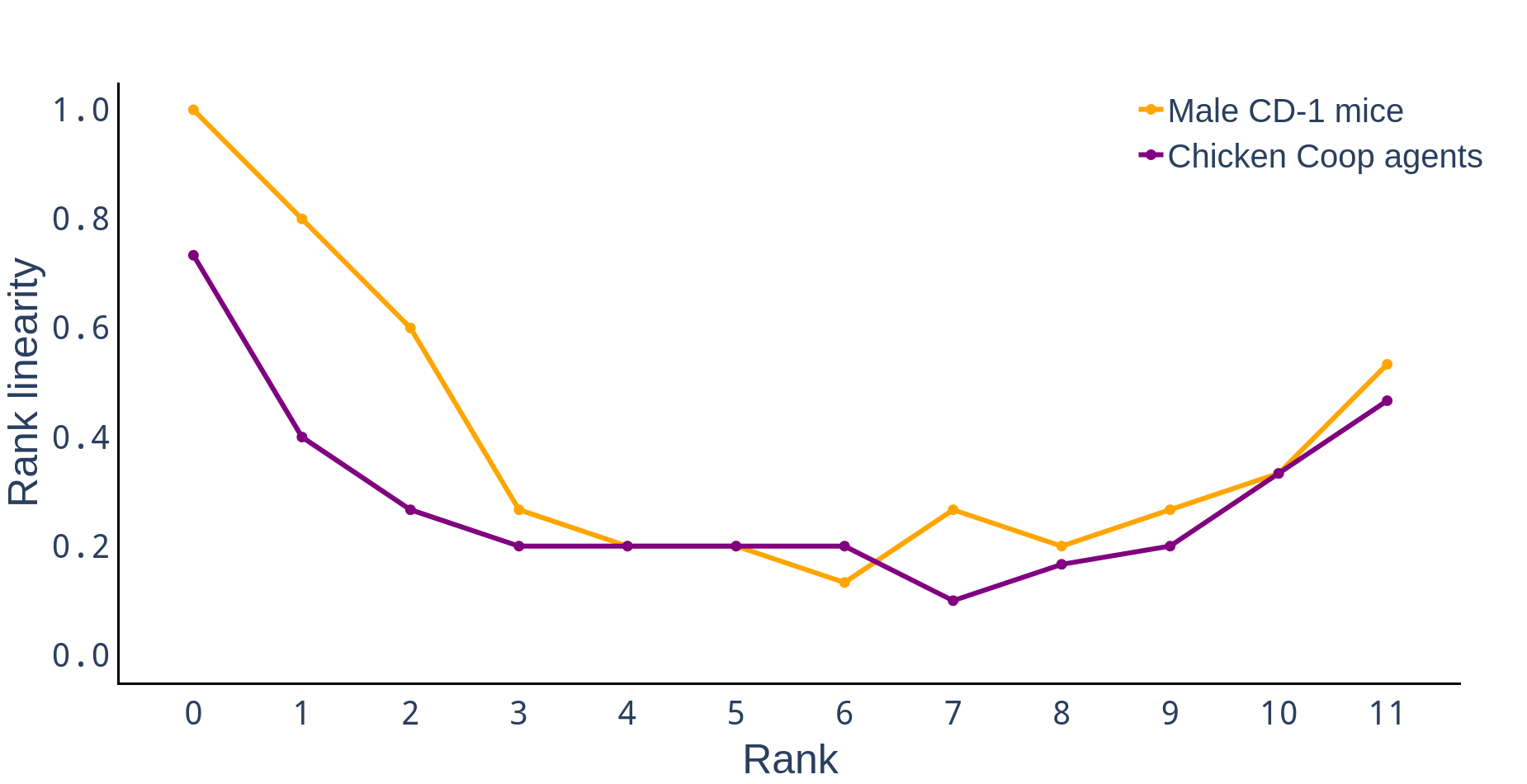}
    \caption{The rank linearities of 12 different ranks, in both Chicken Coop agents (purple) and male CD-1 mice (orange)\citep{williamson2017social}. In both experiments, the top and bottom ranks were the least likely to be occupied by an intransitive component.}
    \label{fig:non_linearity}
\end{figure}

Inspired by the geometric study of intransitive policies in \citet{czarnecki2020spinning}, we visually compare the occurrence of intransitive components (cycles) between Chicken Coop populations and mice populations, shown in \cref{fig:non_linearity}. In \citet{williamson2017social}, 20 captive populations of 12 male CD-1 mice each were observed for aggressive behavior for 22 days.\footnote{The data was conveniently aggregated by \citet{strauss2022domarchive}}. We compare these results to those from $L = 30$ populations of $N = 12$ Chicken Coop agents. In order to find the maximum similarity between the natural and artificial populations, we use a learning rate of $3 \times 10^{-5}$. We make this comparison by using a metric we define here as \textit{rank linearity}\footnote{The metric of rank linearity is similar to a metric called \textit{dominance certainty}, introduced in \citet{mccowan2022certainty} and measured in rhesus macaque breeding groups.}, which expresses how likely each dominance rank is to be occupied by a single agent, rather than a $K$-way tie. For a single population, we can describe rank linearity as a boolean; for example, in the third population from the top in \cref{fig:samples_dh}, the top rank (rank 0) and the two bottom ranks (ranks 4 and 5) are occupied by agents 4, 0 and 2 respectively. Therefore, we consider ranks 0, 4 and 5 as linear. The remaining ranks are 1, 2 and 3, and they cannot be directly correlated with any agent, because agents 1, 3 and 5 are in a three-way tie. Therefore, we consider ranks 1, 2 and 3 as non-linear. For sets of populations, we define rank linearity as a number in $[0, 1]$ which is the probability of a rank being linear in a sample population.

\subsection{Observation ablation}
\label{subsec:ablation}

\citet{tomlinson2002synthetic} proposes that an agent's perception of its opponent agent is a necessary condition for the formation of social relationships between agents.  In this subsection we show a causal relationship between the ability of agents to identify their opponents and their tendency to form dominance hierarchies.

In the Chicken Coop environment, each agent observes the identity of its opponent as an index number between $0$ and $N-1$. We add gradually-rising amounts of random noise to the agents' observation, up to and including its complete obfuscation, and measure the resulting rapport in the population.

\textbf{Methodology:} We define an agent's \textit{opponent perception accuracy} (OPA) as a number between $0$ and $1$. For each agent in each episode, the probability that its observation of its opponent is replaced with a random index number is $(1 - OPA)$. For each OPA value in $\{0, 0.1, 0.2, \ldots, 1\}$ we train $L = 10$ different populations of $N = 6$ Chicken Coop agents at a learning rate of $2 \times 10^{-6}$ and measure their rapport.

\begin{figure}[t]
    \includegraphics[width=\linewidth]{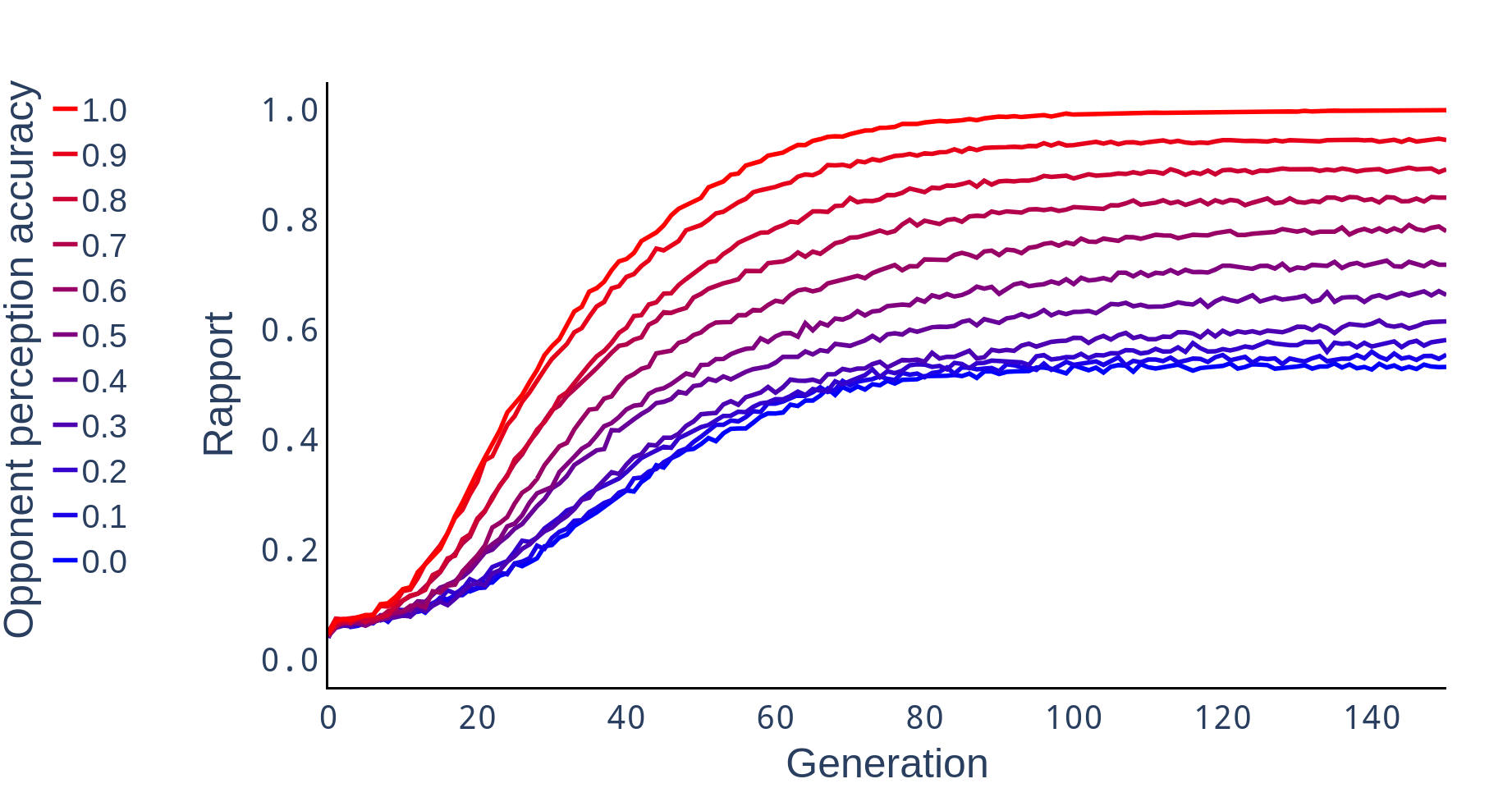}
    \caption{The gradual rise of rapport in Chicken Coop populations, differentiated by how accurately the agents identify their opponents. The bluer lines represent bigger amounts of random noise added to the agents' observation signal.}
    \label{fig:ablate_observation_rapport}
\end{figure}

In \cref{fig:ablate_observation_rapport} we show a positive correlation between the agents' opponent perception accuracy and their rapport, and by extension, their tendency to form dominance hierarchies. 

\subsection{Transmitting dominance hierarchies to new populations}
\label{subsec:transmitting}

In previous subsections we regarded Chicken Coop agents as the focal objects of study, while we considered the dominance hierarchies that they form as a pattern in their behavior. In this subsection we propose flipping this perspective on its head: We consider dominance hierarchies as the focal object while the agents are used as vessels that communicate the dominance hierarchy that they learned to a new population, promoting a cultural evolution paradigm in artificial intelligence \citep{cgi2022learning, tomlinson2009commemorate}.

This transmission of dominance hierarchies to new populations is made possible by two properties of dominance hierarchies:

\begin{enumerate}
    \item
        The knowledge of each dominance hierarchy is replicated over all the agents that participate in it, with each agent having knowledge of a part of that dominance hierarchy in its neural network.
    \item
        Each agent learns to consistently play $hawk$ with all its subordinates, and this tendency serves as a punishing mechanism \citep{axelrod1986evolutionary}, or a sanction. Any errant attempt by a subordinate to play $hawk$ against a dominant would usually result in a $hawk / hawk$ outcome, with both agents receiving the lowest possible reward of $P$. All non-bottom agents effectively act as independent enforcers of their dominance hierarchy, punishing any agent that disobeys it. 
\end{enumerate}

\textbf{Methodology:} We train $L = 10$ different populations of $N = 6$ Chicken Coop agents, similarly to \cref{subsec:emergence}, until they converge to dominance hierarchies. We call these \textit{experienced populations}. For each of these $L$ populations, we perform the following procedure $M = 30$ times: for each $K \in \{0, 1, 2, \ldots, N - 2\} $ we choose a random sample of $K$ of those agents to be our \textit{experienced agents}. We transplant the $K$ experienced agents into a new population with $N-K$ \textit{naive agents} \citep{duenez2021discrimination} that have not been trained. Finally, we train the naive agents in the heterogeneous population at a learning rate of $2 \times 10^{-5}$ and observe their dominance behavior. 

We label the indices of the experienced agents $I_e$, and the indices of the naive agents $I_n$.

To measure how close each new dominance hierarchy is to its original dominance hierarchy, we define a \textit{dominance hierarchy distance} (DHD) as the portion of dominance relationships that have the same polarity in both dominance hierarchies:

\begin{equation}
    \text{DHD}(\mathcal{H}, \mathcal{H}') = 1 - \frac{2 \cdot |E(\mathcal{H}) \cap E(\mathcal{H}')|}{N(N-1)}
\end{equation}

Where $E(\mathcal{H})$ represents the set of directed edges in the dominance hierarchy as ordered pairs. 

To restrict our measurement only to dominance relationships between pairs of naive agents, we define a \textit{restricted dominance hierarchy distance} (RDHD), which is a distance measure similar to DHD, except that only dominance relationships between a specified set $I$ of agent indices are sampled for their polarity:

\begin{equation}
    \text{RDHD}(\mathcal{H}, \mathcal{H}', I) = 1 - \frac{2 \cdot |E(\mathcal{H}) \cap E(\mathcal{H}') \cap (I \times I)|}{|I|(|I|-1)}
\end{equation}

Finally, we define the \textit{dominance hierarchy transmission fidelity} as the complement to RDHD, where the distance is restricted to naive agents only:

\begin{definition}[Dominance hierarchy transmission fidelity]
    The \textit{dominance hierarchy transmission fidelity} (DHTF) between two dominance hierarchies $\mathcal{H}$ and $\mathcal{H}'$, with a common set of naive agents indexed by $I_n$, is a measure of how similar the dominance relationships are between the naive agents in the two different hierarchies:
    \begin{align}
        \text{DHTF}(\mathcal{H}, \mathcal{H}') &= 1 - \text{RDHD}(\mathcal{H}, \mathcal{H}', I_n) \\
        &= \frac{2 \cdot |E(\mathcal{H}) \cap E(\mathcal{H}') \cap (I_n \times I_n)|}{|I_n|(|I_n|-1)}
    \end{align}
\end{definition}

\begin{figure}[t]
    \includegraphics[width=\linewidth]{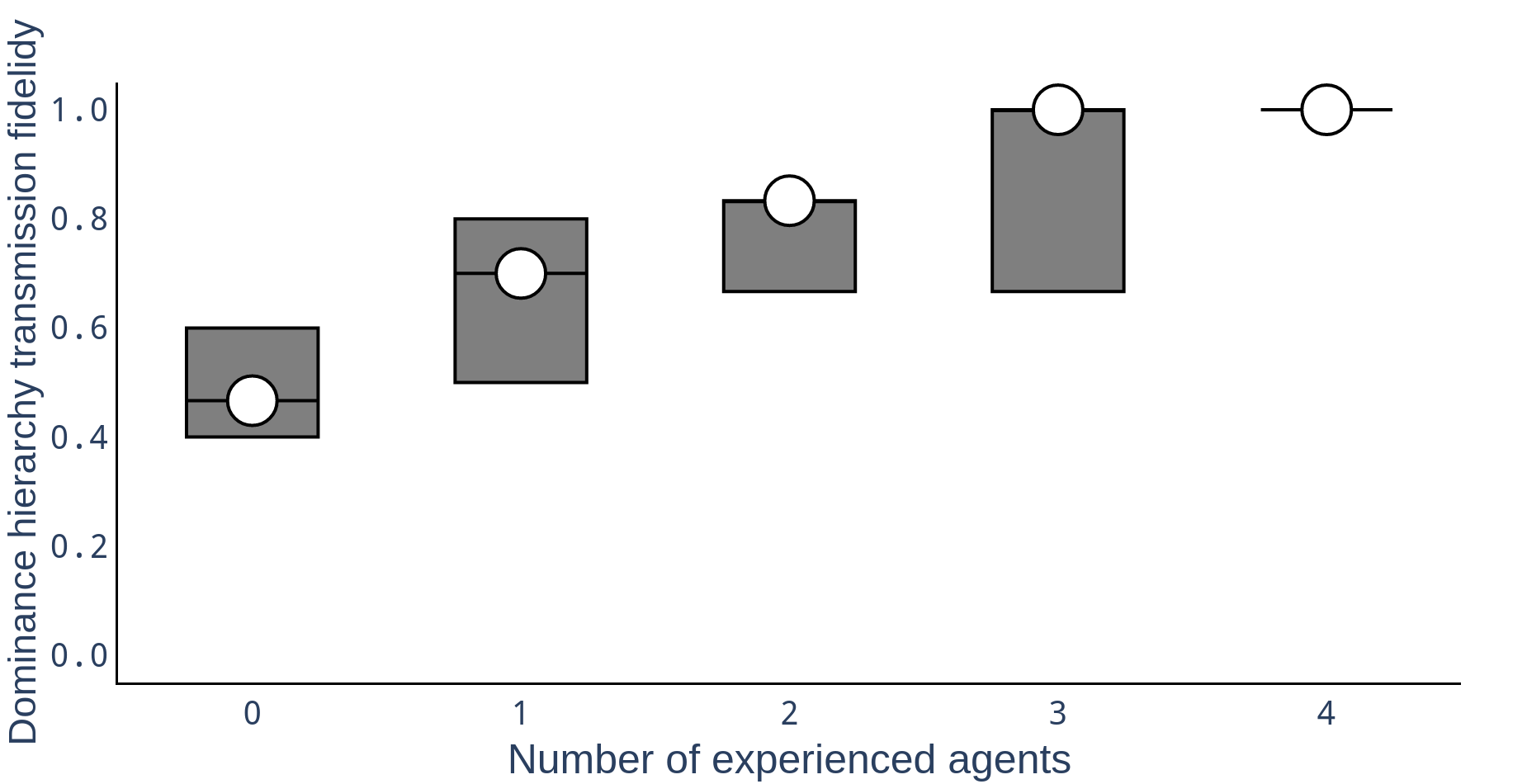}
    \caption{The dominance hierarchy transmission fidelity (DHTF) of 300 different groups of experienced agents transplanted into 300 different naive populations. Higher values mean that the naive agents learned a dominance hierarchy similar to the one in the original experienced population. Circles indicate median DHTF while boxes indicate the two middle quartiles.}
    \label{fig:transmit}
\end{figure}

In \cref{fig:transmit} we show the distribution of the dominance hierarchy transmission fidelities for different values of $K$. A mere $K = 2$ experienced agents are able to teach $4$ naive agents their original dominance hierarchy with $80\%$ median fidelity.

This transmission experiment may be repeated with the naive agents functioning as experienced agents in a new, third population, and so on to an indefinite string of populations, untethering the concept of dominance hierarchies from dependence on any specific host agents.

\section{Discussion and future work}

This paper thoroughly investigates dominance hierarchies in the context of multiagent RL. It first provides a formal definition and a game-theoretical representation of dominance hierarchies. It then introduces the phenomenon of dominance hierarchies in its most simplified form using the Chicken Coop environment, stripping away many of the complexities found in dominance hierarchies in biological life. Finally, it demonstrates how populations of RL agents can invent, learn, enforce, and transmit dominance hierarchies.

Concerning the empirical evaluation, this research was conducted using PPO agents with a single learning regime. This intentional decision was made, in order to mitigate any confounding effects from using a variety of learning mechanisms. However, future research could explore how different RL algorithms learn and adapt dominance hierarchies, as well as the effects of different learning parameters on the emergence and stability of these hierarchies.

Preliminary experiments with higher learning rates result in unstable dominance hierarchies, where agents seem to converge strongly to a certain rank for some number of generations, only to deviate from it and converge to a different rank. This is another unique behavior that exists in natural environments, where it is known as \textit{rank change} \citep{samuels1987continuity, strauss2022dynamics} or \textit{dynamic stability} \citep{chase2022networks}.

We would like to suggest several extensions to the Chicken Coop environment that may provide additional insights into MARL systems:

\begin{enumerate}
    \item \textbf{The Handicap Principle \citep{zahavi1999handicap}:} Enhancing the model with a cost for being dominant. In this environment, an agent always gets more reward by being at a high rank rather than at a low rank. In animal and human societies, a high rank carries sacrifices and risks that may outweigh the benefits of the high rank.
    \item \textbf{Multiple Hierarchies:} In the Chicken Coop environment, all the agents participate in one dominance hierarchy that includes all of the agents. In animal and human societies, individuals participate in multiple dominance hierarchies for multiple groups of other individuals, holding different ranks in each one.
    \item \textbf{Societies with Mixed Incentives:} In this environment, the agents are fully dedicated to making decisions on how aggressive they'll be to other agents. In animal and human societies, individuals in a group cooperate towards shared goals in addition to their internal dominance struggles.
    \item \textbf{Opponent Shaping (OS):} Algorithms such as LOLA \citep{foerster2017lola} and M-FOS \citep{lu2022mfos} enable agents to take the learning process of the other agents into account in their own policy gradients. When applied to an environment that supports dominance hierarchies, they may promote second-order dominance-seeking strategies, e.g., they may prompt agent $i$ to consider how it can behave as to encourage agent $j$ to place agent $i$ at a high rank in the dominance hierarchy.
\end{enumerate}

We propose that the study of dominance hierarchies between RL agents may be combined with research on large language models (LLMs) and multimodal foundation models (MFMs), which have gained immense popularity due to their robustness in solving a wide variety of problems \citep{xi2023rise-llms,li2023multimodal}. \citet{tufano2024autodev} and similar work \citep{vezhnevets2023generative, wu2023autogen} present an innovative way to use LLMs: instead of sending queries to a single LLM, users converse with a group of LLM-based agents, organized in human-inspired hierarchies such as manager-workers and student-assistant-expert. The back-and-forth interaction between multiple LLM agents is effective at encouraging diversity \citep{liang2023divergent}, factuality, and reasoning capability \citep{du2023factuality}, when compared to a single LLM-based agent. We suggest that the augmenting these LLM-based agents with RL or OS algorithms, specifically to maintain the social relationships between the agents, may result in hierarchies that are similar to those observed in biological life.

When humans work on problems as a group, we balance in-group intrigues with external pressures to provide good results. We suggest that this interplay between individual needs and group needs may play a crucial role in the success of our collective intelligence. We hypothesize that MFM-based agents operating under similar conditions may make real-world decisions in a way that is more interpretable and corrigible, as human operators may recognize that the agents' decision process reflects their own.

\begin{acks}

We thank our colleagues for their advice and support in writing this paper: David Aha, Ivan Chase, Edgar Duéñez-Guzmán, Errol King, Joel Leibo, Olof Leimar, David Manheim, Markov, Georg Ostrovski, Jérémy Perret, Saul Pwanson, Venkatesh Rao and Eli Strauss.

This research was funded by The Association For Long Term Existence And Resilience (ALTER).

\end{acks}



\bibliographystyle{ACM-Reference-Format} 
\bibliography{bibliography}


\end{document}